\documentclass[useAMS, usenatbib, usegraphicx]{mn2e}
\usepackage[totalwidth=500pt, totalheight=707pt]{geometry}
\usepackage{times}

\newcommand{\rtt}{\ensuremath{R_{23}}}
\newcommand{\ott}{\ensuremath{O_{32}}}
\newcommand{\OH}{\ensuremath{\log({\rm O/H})}}
\newcommand{\HII}{\mbox{H\,{\sc ii}}}
\newcommand{\OIII}{[\mbox{O\,{\sc iii}}]}
\newcommand{\OII}{[\mbox{O\,{\sc ii}}]}

\newcommand{\NII}{[\mbox{N\,{\sc ii}}]}
\newcommand{\SII}{[\mbox{S\,{\sc ii}}]}

\title[Chemical Evolution of LSB Galaxies]{Oxygen Abundances and Chemical Evolution in Low Surface Brightness Galaxies}

\author[Kuzio de Naray, McGaugh \& de Blok]
 {Rachel Kuzio de Naray,$^1$\thanks{Email:  kuzio@astro.umd.edu (RKD); ssm@astro.umd.edu (SSM); Erwin.deBlok@astro.cf.ac.uk (EdB)} Stacy S. McGaugh,$^1$\footnotemark[1] and W.J.G. de Blok$^2$\footnotemark[1]\\
$^1$Department of Astronomy, University of Maryland, College Park, MD 20742-2421 USA\\
$^2$Department of Physics and Astronomy, Cardiff University, 5, The Parade, Cardiff, CF24 3YB, United Kingdom}

\begin{document}

\date{Accepted 2 September 2004    }

\pagerange{\pageref{firstpage}--\pageref{lastpage}} \pubyear{year}

\maketitle

\label{firstpage}

\begin{abstract}
We report the oxygen abundances of the \HII\ regions of a sample of low surface brightness (LSB) galaxies.  We provide analytic functions describing the McGaugh (1991) calibration of the $R_{23}$ method.  We use this and the
equivalent width (EW) method to determine oxygen abundances, and also
make direct estimates in a few cases where
the temperature sensitive \OIII\ $\lambda$4363 line 
is available.  We find LSB galaxies to be metal poor, consistent with the L-Z relation of other galaxies.  The large gas mass fractions of these objects provide an
interesting test of chemical evolution models.  We find no obvious deviation from the closed-box model of galactic chemical evolution.  Based on our abundance and gas mass fraction measurements, we infer that LSB galaxies are not fundamentally different than other galaxy types but are perhaps at an early stage of evolution.
\end{abstract}


\section{Introduction}
Low surface brightness (LSB) galaxies are normal sized disk galaxies with peak surface brightnesses of $\mu_{o}$ $\geq$ 23$B$ mag arcsec$^{-2}$.  The sky background makes it very difficult to study the weak signal from their stellar continuum. Emission lines from bright \HII\ regions embedded within these galaxies provide
more practical probes.  Prominent among these are the strong lines of oxygen from which the abundance of that element can be deduced.  
As a primary product of Type II supernovae and
the most abundant element after hydrogen and helium, oxygen 
provides a useful measure of the degree of chemical evolution of a galaxy.

LSB galaxies tend to have gas fractions that are systematically larger than well-studied bright galaxies \citep{McGaugh97,Schombert}.  Since gas fraction is the primary variable in theories of chemical evolution, LSB galaxies afford the opportunity to test models like closed-box evolution in an important regime.  This also provides a
check on the luminosity-metallicity (L-Z) relation determined for other galaxy types.   

In this paper we study the optical emission line spectra from \HII\ regions in a number of LSB galaxies.  In Section 2 we discuss the observations and data reduction.  In Section 3 we compare our spectra to diagnostic diagrams.  Oxygen abundances are calculated in Section 4 and discussed in Section 5.  The luminosity-metallicity and mass-metallicity relations are discussed in Section 6, followed by the gas mass fraction in Section 7.  Conclusions are presented in Section 8.

\section{Observations and Data Reduction}
The William Herschel Telescope at La Palma was used to obtain optical spectra of \HII\ regions in six low surface brightness (LSB) galaxies during the nights of 1995 January 4-6.  A 2.4 arcsec slit with 7 \AA\ resolution was used to take blue spectra spanning 3650-5250 \AA\ and red spectra spanning 5700-7300 \AA.  Five standard stars and CuNe and CuAr calibration lamps were also observed for flux and wavelength calibration, respectively.

The LSB galaxies observed were selected for containing reasonably bright, low metallicity \HII\ regions (as determined by \citealt{McGaugh94}).  The slit was placed along each galaxy so its position would maximize the number of \HII\ regions observed during each exposure.  There were two slit positions for UGC 12695.  Individual exposures were 900 seconds, and multiple exposures were taken to increase the signal-to-noise and reject cosmic rays.  Cumulative exposure times ranged from 7200 s to 20,700 s.  Table 1 lists the galaxies and number of \HII\ regions observed.  Images of the six galaxies observed are shown in Figs. 1-6.  North is up and East is to the left in the images.  \HII\ regions with oxygen abundance measurements are labeled in each image.  Further discussion of the locations of the \HII\ regions can be found in McGaugh (1992), and more images are published in McGaugh, Schombert \& Bothun (1995).
\begin{table}
 \caption{Summary of Observations.}
 \begin{tabular}{@{}lcc}
 \hline
  Galaxy  & No. of \HII\ Regions & Total No. Exposures \\
 \hline
  F563-1 &3 &23\\
  F571-5 &1 &9\\
  UGC 1230 &5 &20\\
  UGC 5005 &2 &14\\
  UGC 9024 &2 &19\\
  UGC 12695 (1) &2 &8\\
  UGC 12695 (2) &3 &9\\
 \hline
\end{tabular}

\medskip
Two slit positions were used for UGC 12695.\\
Individual exposures were 900 seconds.
\end{table}

\begin{figure}
 \includegraphics[width=84mm]{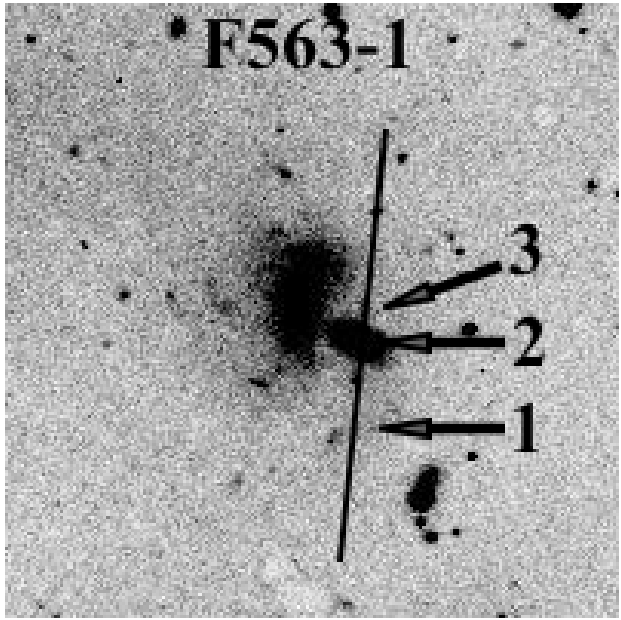}
 \caption{$R$-band image of F563-1.  North is up and East to the left.
 The observed \HII\ regions are labeled along the slit.}
\end{figure}

\begin{figure}
 \includegraphics[width=84mm]{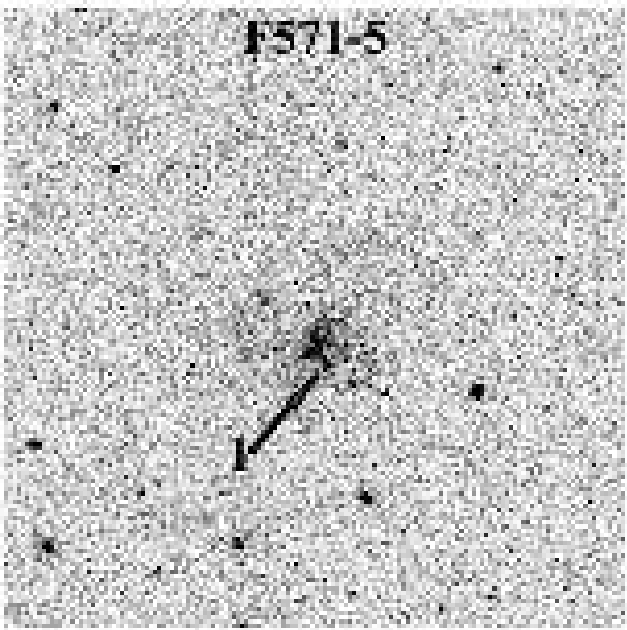}
 \caption{$V$-band image of F571-5.  The observed \HII\ region is labeled.}
\end{figure}

\begin{figure}
 \includegraphics[width=84mm]{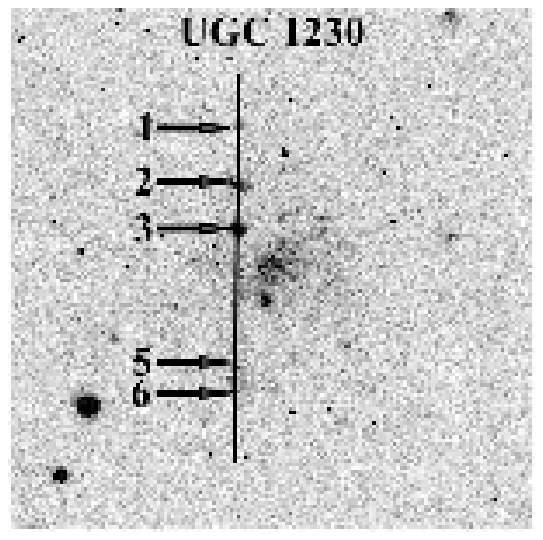}
 \caption{H$\alpha$ image of UGC 1230.  Observed \HII\ regions are labeled along the slit.}
\end{figure}

\begin{figure}
 \includegraphics[width=84mm]{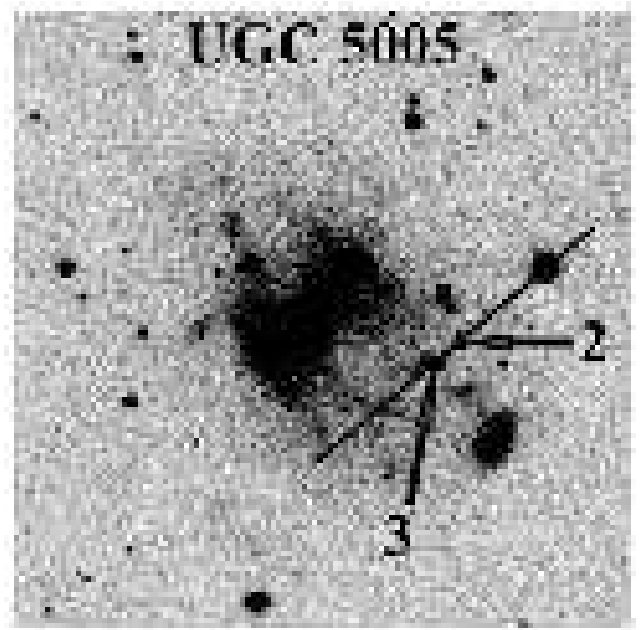}
 \caption{$R$-band image of UGC 5005.  The observed \HII\ regions are labeled along the slit.}
\end{figure}

\begin{figure}
 \includegraphics[width=84mm]{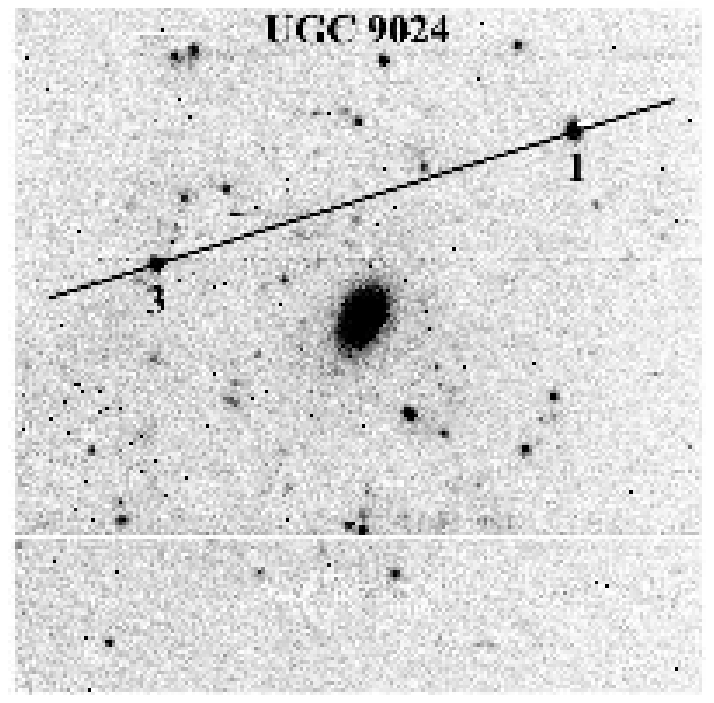}
 \caption{H$\alpha$ image of UGC 9024.  Observed \HII\ regions are labeled along the slit.}
\end{figure}

\begin{figure}
 \includegraphics[width=84mm]{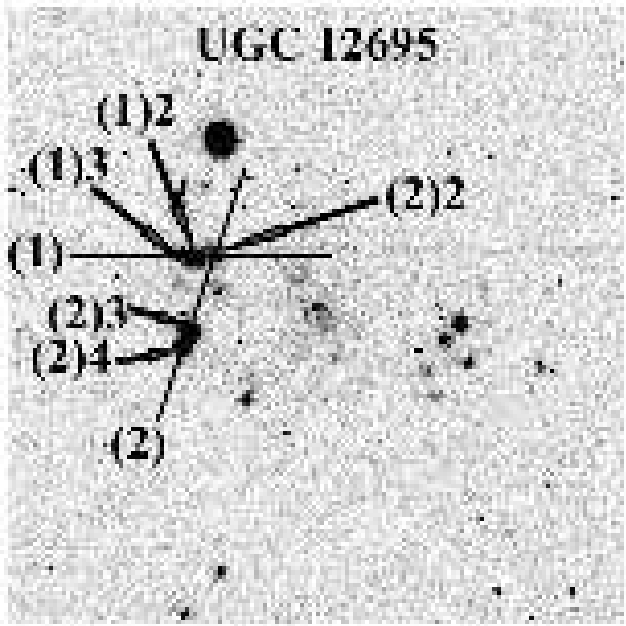}
 \caption{H$\alpha$ image of UGC 12695.  Observed \HII\ regions for both slit positions are labeled.  Slit position (1) runs East-West; slit position (2) runs roughly North-South.}
\end{figure}

Standard data reduction techniques were carried out using IRAF \footnote{The Image Reduction and Analysis Facility (IRAF) is distributed by the Association of Universities for Research in Astronomy, Inc., under contract to the National Science Foundation.}.  First, the average bias frame was subtracted.  Next, sky flats were used to ensure uniform illumination along the slit and tungsten flats were used to flatten the observations in the spectral direction.  Wavelength calibration of the observations was achieved by identifying spectral lines in the CuNe and CuAr calibration spectra.  The \HII\ region spectra were flux calibrated using the standard star observations.  Lastly, the individual \HII\ region spectra were extracted from the galaxy frames and saved as one-dimensional spectra.  The spectra from multiple observations were co-added.  Fig. 7 shows an example of an \HII\ region spectrum with important nebular lines identified.  It should be noted that this is one of the better spectra taken and is not typical of the entire sample.
\begin{figure}
 \includegraphics[width=84mm]{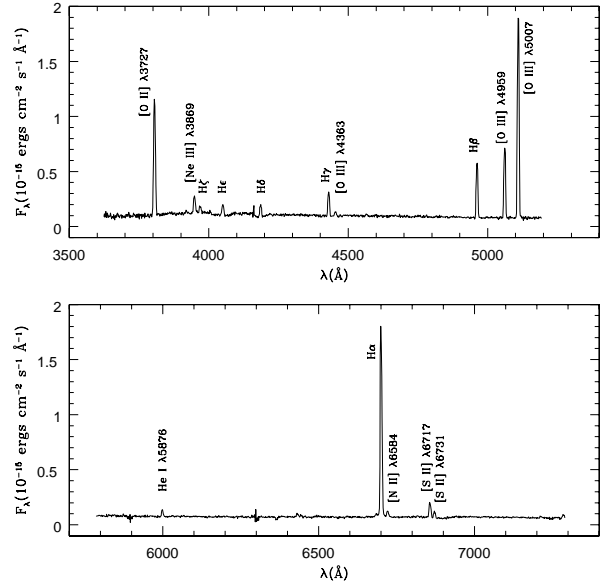}
 \caption{Spectrum of HII region 2 in UGC 12695 slit position 1.  Important nebular lines are labeled.}
\end{figure}

There is no overlap between the blue and red ends of the spectra.  Based on previous experience with the instrument, we anticipated no difficulty in reconciling the calibrations of the red and blue spectra with standard reduction procedures.  While this is often the case, we found that the flux scales are not always consistent in that they have a mismatch between continuum levels or give unphysically low H$\alpha$/H$\beta$ ratios ($\la 2$).  This difference between arms is inconsistent among different galaxies and sometimes even between \HII\ regions of the same galaxy.  We therefore have no confidence in the absolute flux calibration of the data, nor can we confidently connect the flux scales of the red and blue spectra.  However, the line ratios within each spectrum (e.g., \OIII/H$\beta$ and \NII/H$\alpha$) can be measured even if those between red and blue spectra (e.g., H$\alpha$/H$\beta$ and \NII/\OII) cannot.  We therefore restrict our analysis to line ratios and equivalent widths from within each spectrum.  The important abundance sensitive lines are all in the blue spectra, so this is not a great impediment.  The most serious problem which arises is that the reddening cannot be estimated from the H$\alpha$/H$\beta$ ratio, though in many cases single spectra containing both these lines already exist (McGaugh 1994).  We list in Table 2 the observed blue line fluxes relative to H$\beta$ and in Table 3, the observed red line fluxes relative to H$\alpha$.

\section{Diagnostic Diagrams}
Considering the difficulties with reconciling the flux scale between red and blue spectra, it is worth making some checks to confirm that line ratios within each of the red and blue spectra do in fact make sense.  This can be done by comparison to independent data (McGaugh 1994) and with diagnostic diagrams a la Veilleux \& Osterbrock (1987) (Fig. 8). The line in Fig. 8 is the relation for \HII\ regions from Baldwin, Phillips \& Terlevich (1981).  In this plot, AGN fall in the upper right and \HII\ regions along the line.
Fig. 8 confirms that our spectra have line ratios consistent with those of \HII\ regions.

Having confirmed that we have \HII\ region spectra, we now
compare the current data to the data of McGaugh (1994).  In Fig. 9 we plot the ratio of the \OII\ and \OIII\ lines of the current sample to the McGaugh (1994) sample against the flux of the H$\beta$ line from McGaugh (1994).    If the same regions have been
observed then the ratios should be the same. It is clear from the figure that, with the exception of two discrepant \HII\ regions in the \OII\ line, the same

\onecolumn
\begin{table*}
 \centering
 \includegraphics[angle=180, width=180mm]{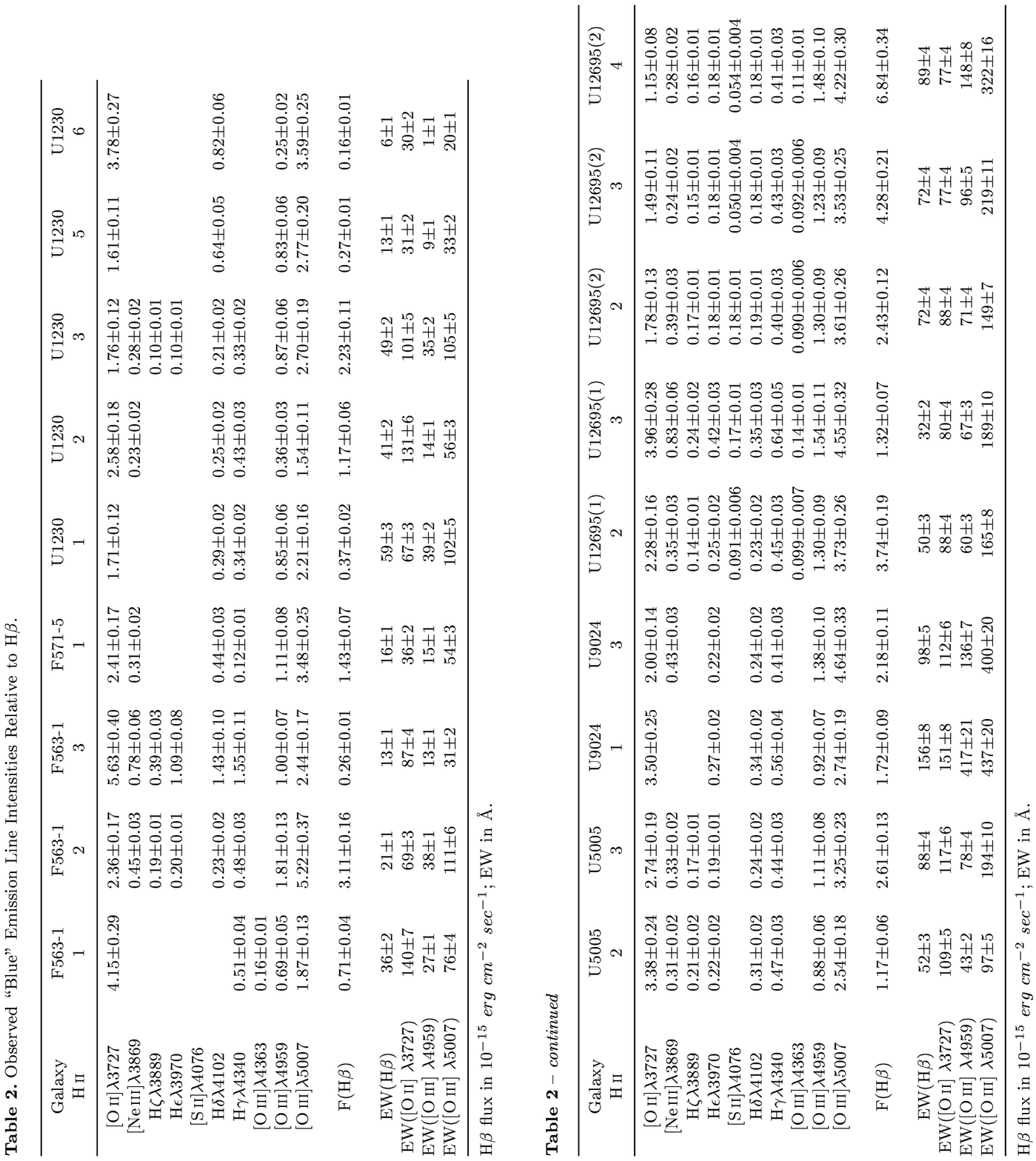}
\end{table*}

\begin{table*}
 \centering
 \includegraphics[angle=180, width=180mm]{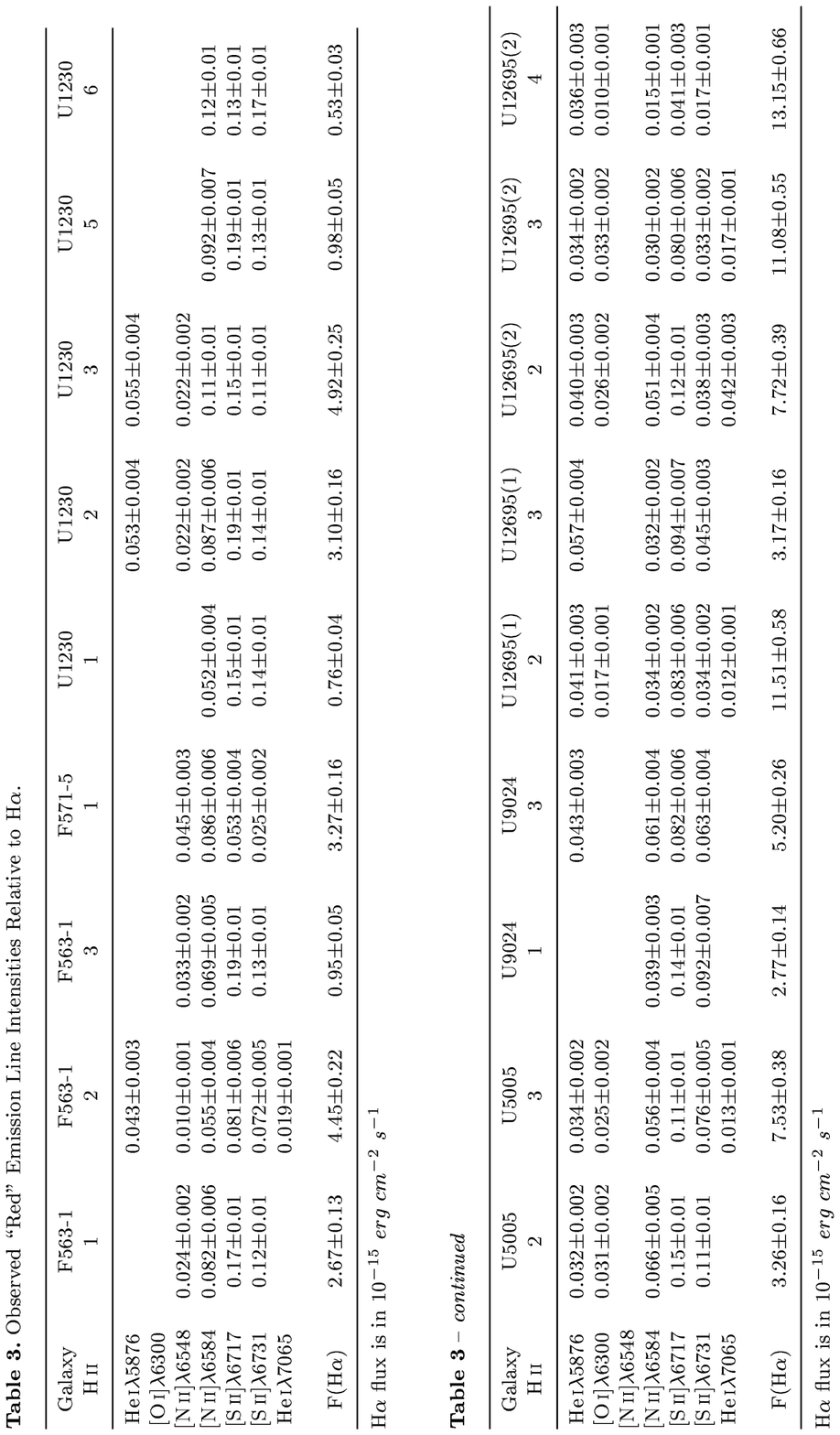}
\end{table*}

\twocolumn

 \HII\ regions have been observed on both occasions.  The two outliers are UGC 1230 \#6 and UGC 12695(1) \#3. UGC 1230 \#6 is a faint \HII\ region and the uncertainties in the original McGaugh (1994) data are large.  The current measurement of the H$\beta$ flux is much less than the measurement of McGaugh (1994). The slit may have missed the majority of this \HII\ region during the more recent observation.  All of the \HII\ regions for both slit positions and both observations of UGC 12695 are large and bright.  Because the \HII\ regions are large, it is possible that they may resolve into smaller ones.  Either the slit fell across a different \HII\ region entirely, or a different part of the same \HII\ region during the two observations.  For these reasons, we exclude \HII\ regions UGC1230 \#6 and UGC 12695(1) \#3 from further analysis.  Together, Figs. 8 \& 9 demonstrate for the rest of the \HII\ regions that even though we cannot intercompare lines from red and blue spectra, we can construct valid line ratios from within each.

\begin{figure}
 \includegraphics[width=84mm]{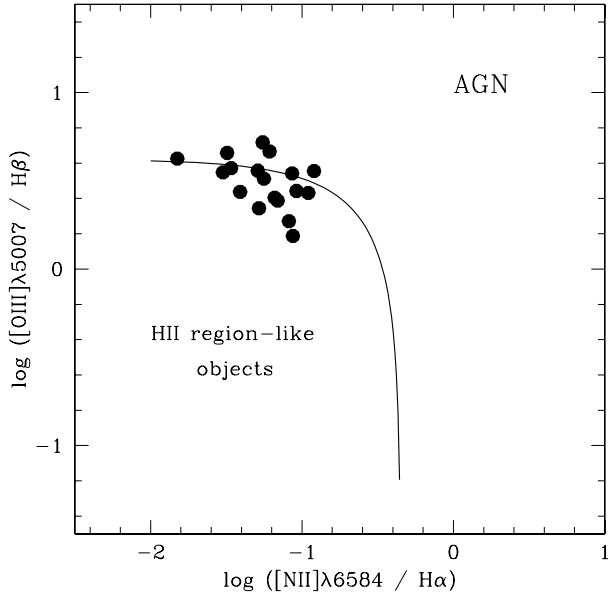}
 \caption{Diagnostic diagram for the \HII\ regions of the current sample.  \HII\ regions should fall along the line from Baldwin, Phillips \& Terlevich (1981).}
\end{figure}

\begin{figure}
 \includegraphics[width=84mm]{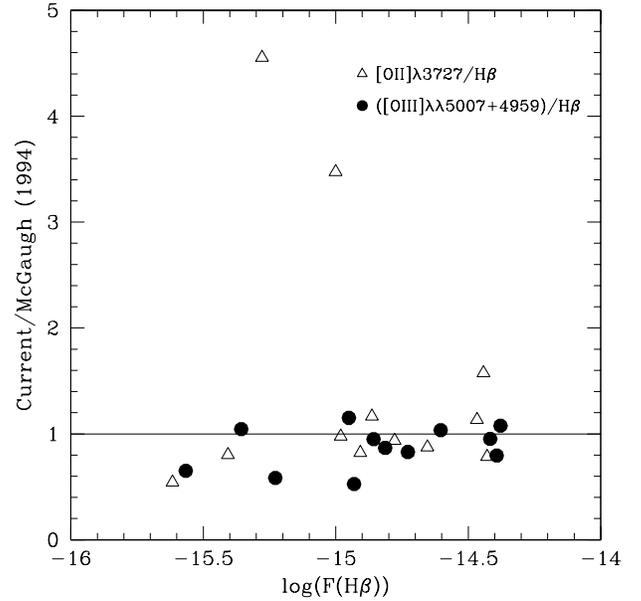}
 \caption{Comparison of oxygen line ratios for the current sample of \HII\ regions to the observations of McGaugh (1994). To separate the \OII\ and \OIII\ data, a shift of 0.05 dex has been applied to the x-axis for the \OIII\ data.}
\end{figure}

\section{Oxygen Abundance}
\subsection{\OIII\ $\lambda$4363 Direct Abundance Measure}
Standard methods of determining oxygen abundances involve measuring a temperature sensitive line such as \OIII\ $\lambda$4363 (Osterbrock 1989).  Ionic abundances of O$^{++}$ and O$^{+}$ can then be derived using the electron temperature.  We have only 4 \HII\ regions which have both an \OIII\ $\lambda$4363 measurement and a reddening correction available from McGaugh (1994) (see Section 4.2).  We determine the temperature using the low-density limit (as indicated by the density-sensitive \SII\ $\lambda$6717/6731 ratio) of the relation between the \OIII\ ($\lambda$4959 + $\lambda$5007)/$\lambda$4363 ratio and the temperature (Osterbrock 1989).  We then follow the method of Pagel et al. (1992) to determine the total oxygen abundance using both O$^{++}$/H$^{+}$ and O$^{+}$/H$^{+}$.  The abundances based on the \OIII\ $\lambda$4363 line are listed in Table 4.  Despite being applicable to only 4 regions, the abundances determined in this fashion can provide an interesting check on abundances derived from indirect techniques.

\setcounter{table}{3}
\begin{table}
 \caption{Direct Oxygen Abundance Measurements}
 \begin{tabular}{@{}lc}
 \hline
  Galaxy \& region &log (O/H) \\
 \hline
  UGC 12695(1) 2 &$-4.27$$\pm$0.07\\
  UGC 12695(2) 2 &$-4.30$$\pm$0.06\\
  UGC 12695(2) 3 &$-4.35$$\pm$0.06\\
  UGC 12695(2) 4 &$-4.32$$\pm$0.04\\ 
 \hline
\end{tabular}

\end{table}

\subsection{$R_{23}$ Strong Line Method}
\OIII\ $\lambda$4363 is notoriously faint, and for much of the present data set is undetected.  When this is the case, one must rely on alternate methods of determining abundances.  A widely accepted approach is the ``R$_{23}$'' strong line method.  As suggested by \citet{Pagel79}, the oxygen abundance (O/H) can be determined from the sum of the \OII\ $\lambda$3727, \OIII\ $\lambda$4959 and \OIII\ $\lambda$5007 intensities relative to H$\beta$.  

The oxygen line strength forms two distinct branches.  At higher abundances, cooling from the infrared lines is important and the sum of the optical \OII\ and \OIII\ lines is low.  As the abundance decreases, cooling comes increasingly from the \OII\ $\lambda$3727 and \OIII\ $\lambda$$\lambda$4959,5007 lines, and the sum of these lines increases. This continues until oxygen becomes rare, and at this point the sum of the \OII\ and \OIII\ lines begins to once again decrease.  The degeneracy between these branches is broken by the ratio of \NII\ $\lambda$6584 to \OII\ $\lambda$3727, although other line pairs may be used (eg. Alloin et al. 1979; van Zee et al. 1998).  $R_{23}$ is:
\begin{equation}
R_{23}\equiv
\frac{{\rm[\mbox{O\,{\sc ii}}]} \lambda3727+{\rm[\mbox{O\,{\sc iii}}]} \lambda4959+{\rm[\mbox{O\,{\sc iii}}]} \lambda5007}{H\beta}.
\end{equation}
Also frequently used in conjunction with $R_{23}$ is $O_{32}$:
\begin{equation}
O_{32}\equiv
\frac{{\rm[\mbox{O\,{\sc iii}}]} \lambda4959+{\rm[\mbox{O\,{\sc iii}}]} \lambda5007}{{\rm[\mbox{O\,{\sc ii}}]} \lambda3727},
\end{equation}
an indicator of the ionization parameter.

There have been several proposed calibrations of the $R_{23}$ method; we use the calibration of \citet{McGaugh91} as represented by the analytic functions described in Appendix A.  These return the oxygen abundance and ionization parameters once the branch of the $R_{23}$ calibration is specified by the ratio of \NII/\OII.  The pure model fit has been used in the following analysis.  

Due to the limitations imposed by the inability to link together the blue and red spectra, we encountered two problems when applying the $R_{23}$ strong line method.  First, we cannot simply use the ratio of \NII\ $\lambda$6584 to \OII\ $\lambda$3727 to break the branch degeneracy--the lines are on separate arms of the spectrum.  Instead we begin by using the \NII\ $\lambda$6584 to H$\alpha$ ratio (van Zee et al. 1998).  Crudely speaking, a ratio greater than 0.1 indicates the upper branch and less than 0.1, the lower branch.  Then, ignoring our calibration difficulties, we determine the branch with the \NII\ to \OII\ ratio and compare the results.  In all cases except UGC 1230 \#5, the branch determination is the same regardless of the method used.  The results using the \NII\ to H$\alpha$ ratio are also the same when compared to McGaugh (1994).  The exception to this is UGC 1230 \#3 which is designated as a lower branch object by McGaugh (1994) but is determined here to be on the upper branch.  For further analysis, we adopt the lower branch designation for both UGC 1230 \#3 and \#5.  The second problem encountered is that the $R_{23}$ strong line method requires that the line fluxes be corrected for reddening.  We are unable to use the H$\alpha$/H$\beta$ ratio to determine reddening corrections because the two lines fall on separate arms of the spectrum.  We attempted to use H$\beta$/H$\gamma$ to obtain the reddening, but the uncertainties were very large.  We therefore adopted the reddening corrections from McGaugh (1994).  This enabled the $R_{23}$ strong line method to be applied to 9 \HII\ regions.  The results are listed in Table 5.

\subsection{Equivalent Width Method}
There are 7 \HII\ regions for which a reddening correction from McGaugh (1994) is unavailable.  For these regions (as well as the others of the current sample), we use an equivalent width version of the $R_{23}$ method to obtain the abundance.  Kobulnicky \& Phillips (2003) found that in the absence of a reddening correction, equivalent width (EW) ratios are an adequate substitute for reddening-corrected flux ratios of the \OII, \OIII, and H$\beta$ lines in the $R_{23}$ method.  To correct for stellar absorption, a 2 \AA\ correction was made to the EW of H$\beta$ for each \HII\ region  \citep{McCall, Oey93, McGaugh94, KobulPhil}.  Following Kobulnicky \& Phillips (2003), the $R_{23}$ strong line method was then implemented using the measured equivalent widths of the lines rather than the fluxes.  Equations 3 and 4 are the equivalent width versions of $R_{23}$ and $O_{32}$:
\begin{equation}
EW_{R_{23}}\equiv
\frac{EW_{{\rm[\mbox{O\,{\sc ii}}]} \lambda3727}+EW_{{\rm[\mbox{O\,{\sc iii}}]} \lambda4959}+EW_{{\rm[\mbox{O\,{\sc iii}}]} \lambda5007}}{EW_{{H\beta}}}
\end{equation}
\begin{equation}
EW_{O_{32}}\equiv
\frac{EW_{{\rm[\mbox{O\,{\sc iii}}]} \lambda4959}+EW_{{\rm[\mbox{O\,{\sc iii}}]} \lambda5007}}{EW_{{{\rm[\mbox{O\,{\sc ii}}]} \lambda3727}}}.
\end{equation}
  The branch was determined using the ratio of the observed, uncorrected \NII\ $\lambda$6584 to H$\alpha$ flux.  The equivalent widths of H$\beta$ and the \OII\ and \OIII\ lines are listed in Table 2 and the resulting oxygen abundances are in Table 5.  We also apply the equivalent width method to a number of other \HII\ regions of McGaugh (1994); the equivalent widths and corresponding oxygen abundances of those regions are listed in Appendix B.

\begin{table}
 \caption{Oxygen Abundances from Indirect Methods}
 \begin{tabular}{@{}lccc}
 \hline
  Galaxy \& region  &log(O/H)  &log(O/H) &log(O/H)\\
     &(EW)  &($R_{23}$) &(McGaugh 1994) \\
 \hline 
  F563-1 1 &$-3.87$$\pm$0.05 & &  \\
  F563-1 2 &$-3.70$$\pm$0.06 & &   \\
  F563-1 3 &$-3.53$$\pm$0.06 & &  \\
  F571-5 1 &$-4.11$$\pm$0.05 & &   \\
  UGC 1230 1 &$-4.49$$\pm$0.04  &$-4.21$$\pm$0.10 &$-4.17$$\pm$0.30 \\
  UGC 1230 2 &$-4.07$$\pm$0.05  &$-4.11$$\pm$0.07 &$-4.04$$\pm$0.16  \\
  UGC 1230 3 &$-4.20$$\pm$0.04  &$-4.17$$\pm$0.03 &$-4.16$$\pm$0.09 \\
  UGC 1230 5 &$-4.20$$\pm$0.05  &$-4.12$$\pm$0.15 &$-3.66$$\pm$0.47 \\
  UGC 5005 2 &$-4.22$$\pm$0.04 & &  \\
  UGC 5005 3 &$-4.35$$\pm$0.04 & & \\
  UGC 9024 1 &$-4.20$$\pm$0.05 & &$-3.95$$\pm$0.21* \\
  UGC 9024 3 &$-4.17$$\pm$0.05 &$-3.79$$\pm$0.23 &$-3.70$$\pm$0.36 \\
  UGC 12695(1) 2 &$-4.13$$\pm$0.05  &$-3.93$$\pm$0.04 &$-4.02$$\pm$0.09 \\
  UGC 12695(2) 2 &$-4.40$$\pm$0.04  &$-4.04$$\pm$0.09 &$-4.01$$\pm$0.12 \\
  UGC 12695(2) 3 &$-4.30$$\pm$0.05  &$-4.12$$\pm$0.04 &$-4.11$$\pm$0.08 \\
  UGC 12695(2) 4 &$-4.26$$\pm$0.05  &$-4.13$$\pm$0.06 &$-4.14$$\pm$0.08 \\
 \hline
 \end{tabular} 
\medskip 

Column 1 abundances are determined using the equivalent width method.
Column 2 abundances use the $R_{23}$ method \& reddening corrections adopted from McGaugh (1994).
Column 3 abundances are from McGaugh (1994).\\
Errors are due to propagation of measurement uncertainties and do not include
calibration uncertainties of at least 0.1 dex.
$^{*}$ Ambiguous branch determination; lower branch abundance is listed here.
\end{table}

\section{Abundance Results and Discussion}
 
\subsection{\HII\ Region Abundances}
We now have at least 1 and up to 4 measurements of the oxygen abundance for each \HII\ region.  Only the quality of the data should affect the abundances using the $R_{23}$ method, as both the method and calibration are identical.  The EW method and the \OIII\ $\lambda$4363 direct method, however, are distinct and those results should be compared to the $R_{23}$ method.  The abundances derived from the \OIII\ $\lambda$4363 measurements agree with  the EW method abundances within the uncertainties.  As was found by McGaugh (1994), the direct oxygen abundances are somewhat lower than those of the $R_{23}$ method.  In Fig. 10 we plot the $R_{23}$-derived abundances of the current data and the McGaugh (1994) data against the EW-derived abundances.  In almost all cases the $R_{23}$ abundances are higher than the EW abundances.  The abundances of McGaugh (1994) are higher than the abundances of the current data, but agree to within the errors.

\begin{figure}
 \includegraphics[width=84mm]{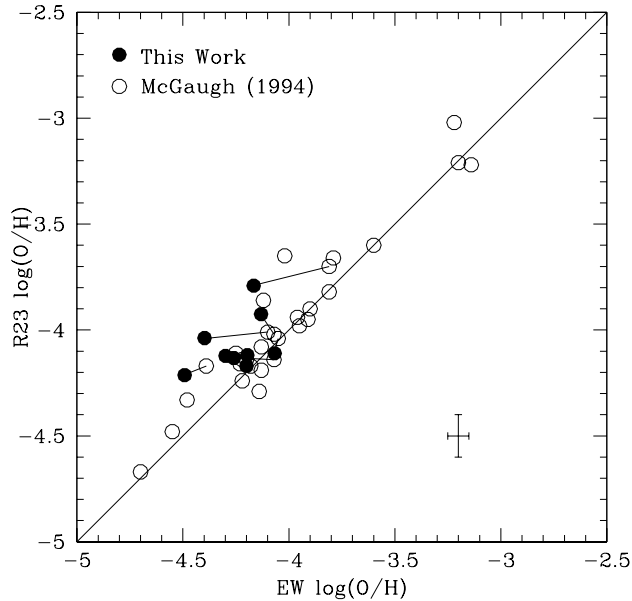}
 \caption{Comparison of oxygen abundances derived using the $R_{23}$ method and the EW method.  Lines connect the regions which are part of both the current sample and the McGaugh (1994) sample.  Crossed lines in the lower right represent typical error bars.}
\end{figure}

\subsection{Mean LSB Galaxy Abundances}
Ideally, we would like to have enough information to determine abundance gradients and use them to establish a characteristic metallicity for each of our galaxies.  This, however, is a difficult challenge in LSB galaxies.  Unlike high surface brightness galaxies where bright \HII\ regions are plentiful, most LSB galaxies lack enough bright \HII\ regions to make an abundance gradient determination (Figs. 1---6).  We therefore choose to take the mean of the abundances of the individual \HII\ regions as representative of the entire galaxy.  One should bear in mind that this mean value may or may not be a good indicator of the true global metallicity.  Confidence in the use of the mean abundance as the overall galaxy abundance seems to be appropriate, though, as de Blok \& van der Hulst (1998) find no abundance gradient, but instead a constant abundance with radius, for 3 LSB galaxies which had enough \HII\ regions distributed over the length of the galaxy to make a gradient study feasible.  It should also be noted that abundances correlate with local surface density as well as with radius (McCall 1982).  By definition, LSB galaxies have lower surface densities and little dynamic range in the surface brightness over which \HII\ regions are observed.  One would thus expect little change in metallicity across an LSB galaxy.  With this in mind, we take the average of the \HII\ region abundances in each galaxy to represent the global abundance of that galaxy.  Table 6 lists both the average EW abundance and the average $R_{23}$ abudance.  The data for our current sample of LSB galaxies has been supplemented with data from McGaugh (1994) and de Blok \& van der Hulst (1998). The McGaugh (1994) data are for 3 \HII\ regions from our galaxy sample that were not observed as part of the current sample. The de Blok \& van der Hulst (1998) data are $R_{23}$ method oxygen abundances for different \HII\ regions of many of the same galaxies as investigated here.

\begin{table*}
\begin{minipage}{120mm}
 \caption{Mean Oxygen Abundances and Gas Mass Fractions }
 \begin{tabular}{@{}lcccccccc}
 \hline
  Galaxy  &$<$log(O/H)$>$ &$<$log(O/H)$>$ & Ref. &$M_{B}$ &(B-V) &$M_{HI}$/$L_{B}$ &$f_{g}$ &Method \\
   &EW &R23 & & & & & &\\
 \hline
  F563-1 &$-3.72$ &$-3.98$ &3  &$-17.33$ &0.65 &2.05 &0.49 &4 \\
  F571-5 &$-4.11$ &$-3.92$ &3 &$-17.14$ &0.34 &1.55 &0.77 &5 \\
  UGC 1230 &$-4.17$ &$-4.09$ &1,2,3 &$-18.33$ &0.52 &1.72 &0.68 &4 \\
  UGC 5005 &$-4.29$ &$-3.96$ &3 & & &1.0* &0.35 &4 \\
  UGC 9024 &$-4.02$ &$-3.65$ &1,2 &$-18.73$ & &0.82 &0.50 &6 \\
  UGC 12695 &$-4.20$ &$-4.06$ &1,2 &$-18.92$ &0.37 &1.28 &0.71 &5 \\
  F415-3 &$-3.96$ &$-3.94$ &2  &$-16.48$ &0.52 &1.71 &0.65 &5\\
  F469-2 &$-4.12$ &$-4.11$ &2 &$-17.39$ &0.43 &0.95 &0.59 &5\\
  F530-3 &$-4.22$ &$-4.14$ &2 &$-18.77$ &0.64 &0.53 &0.26 &5\\
  F561-1 &$-4.12$ &$-3.88$ &2,3 &$-17.83$ &0.59 &0.68 &0.36 &5\\
  F563-V1 &$-4.30$ &$-4.43$ &2 &$-16.36$ &0.56 &0.90 &0.45 &5\\
  F563-V2 &$-4.24$ &$-3.90$ &2,3 &$-18.21$ &0.36 &0.76 &0.61 &5 \\
  F568-6 &$-3.64$ &$-3.36$ &2 &$-21.57$ &0.63 &0.44 &0.23 &5\\
  F577-V1 &$-3.76$ &$-3.60$ &2 &$-18.21$ &0.38 &0.83 &0.61 &5\\
  F611-1 &$-4.34$ &$-4.23$ &2 &$-15.76$ &0.44 &1.07 &0.61 &5\\
  F746-1 &$-3.71$ &$-3.66$ &2 &$-19.42$ &0.65 &0.73 &0.32 &5\\
  UGC 5709 &$-3.21$ &$-3.20$ &2 &$-19.77$ &0.48 &0.52 &0.40 &5\\
  UGC 6151 &$-3.79$ &$-3.94$ &2 &$-17.49$ &0.51 &0.64 &0.42 &5\\
 \hline
\end{tabular}
\medskip

 References for mean R23-derived abundances$-$ 1: Current data;  2: McGaugh (1994);  3: de Blok \& van der Hulst (1998)\\
 Gas Mass Fraction Methods$-$ 4: de Blok \& McGaugh (1998);  5: Uses Eq. 7 of the text and relevant data from McGaugh \& de Blok (1997);  6: McGaugh \& de Blok (1997)\\
 $^{*}$ $M_{HI}$/$L_{R}$ from van der Hulst et al. (1993)
 
\end{minipage}
\end{table*}

\section{Luminosity-Metallicity Relation}
The luminosity-metallicity (L-Z) relation, and mass-metallicity relation, have been important tools for studying galactic chemical evolution.  Relations have been found for, among others, dwarf Irregulars (eg. Lee et al. 2003), late-type galaxies (eg. Pilyugin \& Ferrini 2000) and ellipticals (eg. Brodie \& Huchra 1991).  L-Z relations have also been found using very large survey samples (eg. KISS Melbourne \& Salzer 2002; 2dFGRS Lamareille et al. 2004; SDSS Tremonti et al. 2004).  The correlation between luminosity and metallicity exists over a wide range of luminosities and metallicities, although there are indications that perhaps dwarf galaxies have a shallower L-Z relation and that one L-Z relation is therefore insufficient to describe all galaxy types.  It is interesting to see where LSB galaxies fit in.

Fig. 11 is the L-Z plot for the LSB galaxies.  Also shown are a variety of recent fits to the L-Z relation.  There is some disparity here which may stem in part from the use of different methods and calibrations for estimating abundances.  It appears that the LSB data prefer a steeper relation similar to that describing the 2dFGRS sample.

\begin{figure}
 \includegraphics[width=84mm]{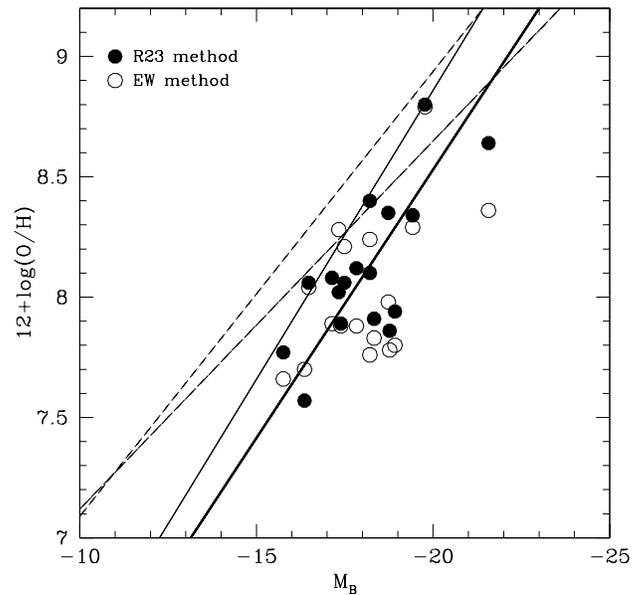}
 \caption{Luminosity-Metallicity plot for LSB galaxies.  The thin solid line is from the KISS sample of Melbourne \& Salzer (2002).  The heavy solid line is from the 2dFGRS sample of Lamareille et al. (2004).  The short-dash line is the Tremonti et al. (2004) SDSS sample, and the long-dash line is for the dIrr sample of Lee et al. (2003).  The majority of data from the Tremonti et al. (2004) fit have $M_{B}$ brighter than $-$18.  The fit is extrapolated to fainter magnitudes in their fig. 4, and further extrapolated here.   }
\end{figure}

Because it is more difficult to estimate mass, mass-metallicity relations are less often computed than L-Z relations.  We have mass estimates for our sample, and present the mass-metallicity plot in Fig. 12.  Also plotted in Fig. 12 is the dIrr sample of Lee et al. (2003) and the results of the mass-metallicity relation determined for SDSS galaxies of Tremonti et al. (2004).  Tremonti et al. (2004) find the correlation between mass and metallicity to be roughly linear between $10^{8.5}$$M_{\sun}$ and $10^{10.5}$$M_{\sun}$ before it begins to flatten out.  The LSB galaxies mainly fall within this linear region, but have considerably lower abundances in the mean.

The calibration of both mass and metallicity is slightly different here than in Tremonti et al. (2004).  The $R_{23}$ calibration used here would give lower (O/H), while the IMF used here is 0.15 dex more massive (Bell et al. 2003).  This would shift the SDSS line down and to the right in Fig. 12.  While the shift in mass is trivial to achieve, that in abundance is more complex because the shape of the calibrations are somewhat different.  For now we ignore these differences as the net difference is not large compared to the range of the plot.  One should bear in mind the caveat that it is
impossible to exclude the possibility that the apparent differences might be reconciled
by a change in calibration (see appendix A).

Accepting the calibration of the mass-metalicity relation as published by Tremonti et al. (2004), the mean abundance of both LSB galaxies and the dIrr sample of Lee et al. (2003) is lower than that of the SDSS galaxies.  This is qualitatively consistent with higher gas fraction galaxies being less evolved at a given current stellar mass.  Certainly it is possible that LSB galaxies simply reside on the low metallicity side of the overall distribution, a result which is consistent with previous work (McGaugh 1994; de Blok \& van der Hulst 1998).  However, the mass-metallicity relation is not well defined below $\log M_{*} = 8.5$, so it is also possible that the relation turns down sharply at this point.

\begin{figure}
 \includegraphics[width=84mm]{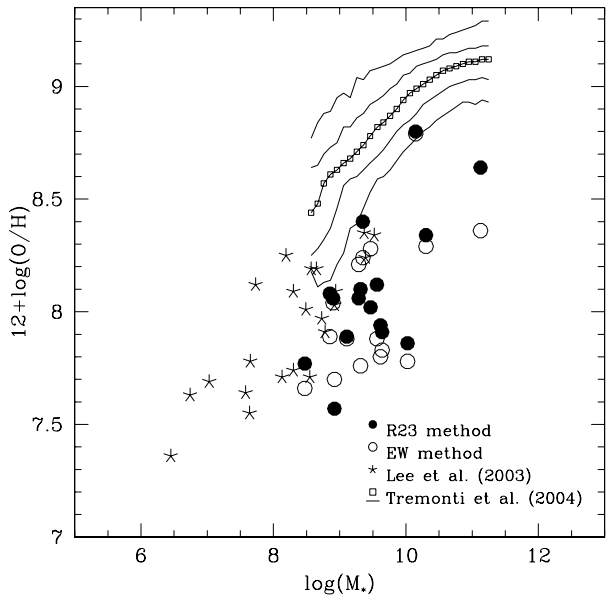}
 \caption{Mass-Metallicity plot for LSB galaxies.  LSB galaxies are represented by circles.  The stars are the dIrr sample from Lee et al. (2003) and the contours are the results of Tremonti et al. (2004) for 53,000 SDSS star-forming galaxies.  The contours enclose 68\% and 95\% of their data. }
\end{figure}

\section{Oxygen Abundance vs. Gas Mass Fraction}
Additional clues to galactic chemical evolution can be gleaned from the gas mass fraction.  The gas mass fraction ($f_{g}$) is a measure of the total mass in gas ($M_{g}$---molecular, atomic, ionized and metals) to the combined total mass in gas and stars ($M_{*}$):
\begin{equation}
f_g=\frac{M_g}{M_g + M_*}.
\end{equation}
Rephrased in terms of observables, the gas mass fractions is
\begin{equation}
f_g=(1 + \frac{\Upsilon_* L}{\eta M_{HI}})^{-1}
\end{equation}
where $\Upsilon_{*}$ is the stellar mass-to-light ratio ($M_{*}$=$\Upsilon_{*}$L), $L$ is the luminosity, $M_{HI}$ is the mass in neutral hydrogen and $\eta$ serves as the equivalent to $\Upsilon_{*}$ ($M_{g}$=$\eta$$M_{HI}$). 

Together, metallicity and the gas mass fraction can be used as a diagnostic for testing different models of the chemical evolution of galaxies.  In the closed-box model, there is neither inflow nor outflow of mass as the initially metal-free gas evolves with time.  Mathematically, this is represented as
\begin{equation}
Z=p\; \ln\left(\frac{1}{f_{g}}\right)
\end{equation}
where Z is the metallicity and p is the yield (eg. Searle \& Sargent 1972; Tinsley 1980; Edmunds 1990).  Following \citet{Lee}, this can be rewritten in a more user-friendly fashion:
\begin{equation}
12 + \log(\rm{O/H}) = 12 + \log(0.196y_{O}) + \log\, \log(1/f_{g})
\end{equation}
where $y_{O}$ is the oxygen yield.  Written this way, the closed-box model predicts a slope of unity; deviations from this slope may be an indication for the inflow or outflow of gas.

Our best estimates of the gas fractions for the entire disk are listed in Table 6.  We give preference to stellar mass estimates from de Blok \& McGaugh (1998).  For galaxies not in that sample, we compute the gas fractions with Eq. 6 using $\eta$=1.4 and compute $\Upsilon_{*}$ from $(B-V)$ using
\begin{equation}
\log(M_{*}/L_{B})=1.737(B-V)-0.787
\end{equation}
\citep{Bell} (their `diet' Salpeter IMF).  We choose this IMF because it is fairly close to optimal with respect to dynamical data (McGaugh 2004).  Finally, if neither of the two previous methods can be applied, we assume the gas fraction in table 1 of McGaugh \& de Blok (1997).  Plotted in Fig. 13 is oxygen abundance versus gas fraction for our data and data for field dIrr from \citet{Lee}.  There is no clear evidence of deviation from the closed-box model evident in this plot.  Our data are consistent with those of \citet{Lee} and \citet{Pagel86} albeit with more scatter.

LSB galaxies are consistent with the yield of p=0.002 determined for other galaxy types (Lee et al. 2003; Pagel 1986).  This is an interesting result for the LSB galaxies as it is often argued that they have shallow potential wells, making it easier for mass loss to occur.  \citet{Garnett} finds that outflows may be important in galaxies with $V_{rot}$$<$50 km $\rm{s}^{-1}$.  All galaxies discussed here have $V_{rot}$$>$50 km $\rm{s}^{-1}$.  The apparent uniformity of the yield is also suggestive of a universal IMF, as favored by dynamical constraints (McGaugh 2004).  However, the abundance data alone leave substantial room to consider other models, such as the bottom-heavy IMF for LSB galaxies suggested by Lee et al. (2004).

\begin{figure}
 \includegraphics[width=84mm]{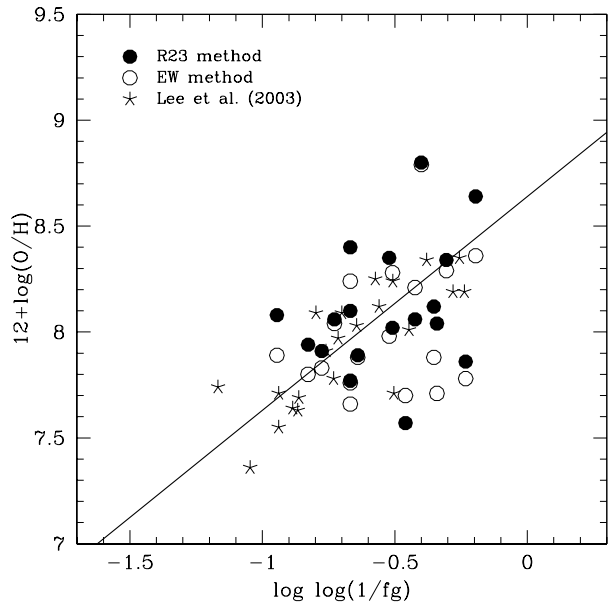}
 \caption{Plot of gas mass fraction versus metallicity.  Circles are the current data set of LSB galaxies.  Stars are data for field dwarf Irregulars from \citet{Lee}.  The solid line indicates the behaviour of the closed box model with a yield of 0.002.}
\end{figure}

\section{Conclusions}
We have studied the optical emission line spectra of \HII\ regions in six LSB galaxies.  We have supplemented this sample with the LSB galaxy data of McGaugh (1994) and de Blok \& van der Hulst (1998).  We have measured oxygen abundances using both the EW method and $R_{23}$ strong line method, and in a few cases directly using the \OIII\ $\lambda$4363 temperature sensitive line.  The oxygen abundances of the LSB galaxies show them to be metal poor, while the gas mass fractions are high.  High gas mass fractions and low metallicities are consistent with LSB galaxies being at an early stage of their evolution. There is no need to suppose that they are fundamentally different from other types of galaxies.  In gross terms, a closed-box model with the same effective yield as for other galaxies appears to be adequate.  There is no evidence in the gas fractions that requires extensive gas loss stemming from blow-out in these systems.  Finally, we have found that LSB galaxies are broadly consistent with the luminosity-metallicity relation of other galaxy types.  

\section*{Acknowledgments}
This research has made use of the NASA/IPAC Extragalactic Database (NED) which is operated by the Jet Propulsion Laboratory, California Institute of Technology, under contract with the National Aeronautics and Space Administration.  The work of SSM is supported in part by NSF grant AST0206078.\\

\appendix

\section{Analytic Expressions for Strong Line Oxygen Abundances}

The \rtt\ method \citep{Pagel79}
is a valuable tool for obtaining oxygen abundance estimates.  
The strong lines upon which it is based are readily accessible observationally.
While the temperature sensitive [O III] $\lambda$4363 line can usually
only be observed in bright, nearby, low metallicity \HII\ regions, the strong
lines are often measurable to cosmologically significant distances.

To facilitate the use of the \rtt\ method for large data sets, we
provide here sets of equations relating the observables \rtt\ and \ott\ to
the oxygen abundance and ionization parameter.
To simplify the appearance of these relations, we employ the following
notation:
\begin{eqnarray}
Z & \equiv & \OH \\
U & \equiv & \log{\bf U} \\
x & \equiv & \log(\rtt) \\
y & \equiv & \log(\ott).
\end{eqnarray}
$Z$ is used here as a shorthand for the logarithmic
oxygen abundance, and should not be
confused with the mass fraction of metals.
{\bf U} is the volume averaged ionization parameter, being basically
the ratio of the number of ionizing photons to the number of particles.

Below we provide functions $Z(x,y)$ and $U(x,y)$ that give the
nebular parameters.
Two similar but distinct calibrations are provided.  The first
is a straight representation of the model grid of \citet{McGaugh91}.
The second attempts a modest correction to this grid to account for
the observed decrease in stellar effective temperatures at high metallicity
by fitting to the metal rich \HII\ region Searle 5 \citep{Shields, Kinkel}.  These relations have been employed to estimate
oxygen abundances in high redshift galaxies by \citet{Lilly} and \citet{Kobulnicky}.

\subsection{Model Calibration}

A functional representation of the $M_u = 60\; M_{\sun}$ model grid
of \citet{McGaugh91} is given here.  The \rtt-O/H relation is double valued,
so we split it into separate expressions for the upper (high-$Z$) and lower
(low-$Z$) branches.  On the lower branch,
\begin{eqnarray}
Z_{\ell}(x,y) = & -4.93  + 4.25x -3.35 \sin(x) \nonumber \\
	& -0.26y - 0.12 \sin(y) \\
U_{\ell}(x,y) = & -2.95  + 0.17x^2  + 1.02y, 
\end{eqnarray}
while on the upper branch,
\begin{eqnarray}
Z_u(x,y) = & -2.65 - 0.91x + 0.12y \sin(x)  \\
U_u(x,y) = & -2.39 - 0.35x + 1.29y - 0.15xy.
\end{eqnarray}
These expressions reproduce the model grid to within 0.1 dex
when the arguments of the trigonometric functions are treated as
being in radians.

Distinguishing which branch is appropriate can be
accomplished with the [N II] $\lambda$6583 line \citep{McGaugh91,McGaugh94}.
Strong (weak) nitrogen emission indicates that the upper (lower) branch
is appropriate.  The dividing line is at $\log({\rm [N\; II]/[O\; II]}) 
\approx -1$.  This can often be judged by eye in raw spectra:
is [N II] prominent compared to H$\alpha$, or is it very weak?
However, there is no precise quantitative dividing line, as the
[N II]/H$\alpha$ ratio depends on a variety of factors. 

The [N II]/[O II] ratio is itself a very good abundance indicator.
In two respects it is superior to \rtt:  (1) it is monotonic rather than
double-valued, and (2), being formed from lines of the same ionization state,
is insensitive to the ionization parameter {\bf U}.  Unfortunately it is rather
difficult to calibrate because of the behavior of N/O at low O/H, and
in practice is very sensitive to the reddening correction because of
the large separation of the lines in wavelength.

\subsection{Semi-Empirical Calibration}

A large variety of \rtt\ calibrations have been suggested since
the method was introduced by Pagel (e.g, see the
compilations of \citealt{McGaugh91, Zaritsky, Kewley}).  Some of these calibrations are empirical,
some are based on models, and others are merely hybrids of previous
calibrations.  The plain fact is that it remains difficult to pin an
absolute calibration on \rtt.

The problem of absolute calibration is particularly acute at
high ($\ga$ solar) metallicity where there is a dearth of independent
data which test the method.  On the one hand, it is clear that \rtt\ is
very sensitive to oxygen abundance, and hence a good {\it relative\/}
abundance indicator.  At and above solar metallicity, slight changes in
O/H cause large ones in \rtt, so the discrimination is very powerful.
However, the power to discriminate between slightly different oxygen
abundances does not extend to knowledge of the absolute abundance.

We provide here a semi-empirical calibration which attempts to
fit the one well known high metallicity point Searle 5.
The original determination \citep{Shields} of the abundance of
Searle 5 is $Z = -2.89$.  \citet{Kinkel} estimate
a lower oxygen abundance of $Z = -3.12$.  The latter determination 
is consistent with the detailed modeling of \citet{Evans}.

Nearly all \rtt\ calibrations in the literature overestimate the lower
abundance measurement,
including the model grid of \citet{McGaugh91} as given in the previous section.
The model grid of \citet{McGaugh91} does give this lower abundance
if one takes $T_{\star} \approx 38,000 K$ as inferred by \citet{Shields} and subsequent workers.  The chief problem seems to be in the
ionizing spectra of metal rich stars which are needed as input to the
photoionization models.  This is a highly nonlinear problem beyond the
scope of this paper.  Nevertheless, we can 
interpolate through the existing model grid to follow the observed drop
in $T_{\star}$ with $Z$.  In effect, we have added an additional term to the
$T_{\star}(M,Z)$ relation discussed by \citet{McGaugh91} to empirically
compensate for the excess hardness of the non-LTE stellar atmospheres
employed in those models.

The result is a semi-empirical calibration which splits the difference
between the \citet{Shields} and \citet{Kinkel} measurements,
yielding $Z = -3.03$ for Searle 5.  This is within 0.1 dex of the
\citet{Kinkel} measurement, which is about what we expect
for the accuracy of this approach.
Polynomial fits to the semi-empirical calibration are
\begin{eqnarray}
Z_{\ell}(x,y) = & -4.944 +0.767 x+0.602 x^2 \nonumber \\
	& -y(0.29+0.332 x -0.331 x^2)
\end{eqnarray}
and
\begin{eqnarray}
Z_u(x,y) = & -2.939 -0.2 x-0.237 x^2-0.305 x^3 \nonumber \\
	& -0.0283 x^4 -y(0.0047 -0.0221 x \\
	& -0.102 x^2 -0.0817 x^3 -0.00717 x^4). \nonumber
\end{eqnarray}
The terms involving $xy$ and $x^4y$ in the expression for $Z_u(x,y)$ have
little impact on the fit but are included for completeness.

Note that {\bf U} and its indicator \ott\ play an important second
parameter role on the {\it lower\/} branch which persists to rather
high abundances ($Z \approx Z_{\sun}$, $x \approx 0.5$).
This effect is not suppressed.  In fact, it is particularly important around
$Z \approx 1/2 Z_{\sun}$, $x > 0.7$.  Many data lie in this region
where other calibrations rely on only a single parameter.  However, at
still higher abundances, the sensitivity of \rtt\ to the oxygen abundance 
becomes so great that it effectively becomes a one parameter sequence. 
If there is any residual second parameter dependence on the upper branch,
it is more likely due to $T_{\star}$ than {\bf U}.

The net result of this modest recalibration of the model grid is a slower
variation of the oxygen abundance with \rtt\ relative to most other
calibrations.  For example, weak \rtt\ lines with $x = 0$ give
$Z = -2.93$, 0.2 dex less than the calibrations of \citet{Edmunds}, McCall et al. (1985) and Zaritsky et al. (1994),
but comparable to that of \citet{Dopita}.

The expressions given here provide a uniform, self-consistent, accurate
{\it relative\/} calibration of the \rtt\ abundance sequence.
It is difficult to estimate the error in the {\it absolute\/} calibration:
we do not know what the intrinsic \rtt-O/H relation actually is.
Our calibration does at least provide a consistent, easy to use method for
estimating abundances over a very large range of parameters,
explicitly incorporating the important effects of ionization.

{\bf An additional note:} Long after this appendix was written, and very late in
the refereeing process, a paper appeared containing more direct estimates
of the oxygen abundance in high abundance \HII\ regions in M51 (Bresolin,
Garnett, \& Kennicutt 2004).  The semi-empirical calibration above, while
lower than most other calibrations and consistent with the one previously
well known case, overestimates the abundances of Bresolin et al. (2004)
by $\sim 0.2$ dex.  The $R_{23}$ calibrations discussed here are still valid
indicators of the relative oxygen abundance, but the results of Bresolin et al.
(2004) highlight the difficulty of obtaining an absolute calibration.  Since the strong line method also tends to overstate lower branch abundances by a comparable amount,
a simple shift might be crudely acceptable.  This would amplify the disparity
between the LSB galaxy abundances determined here and the mass-metallicity relation of Tremonti et al. (2004).  If the results of Bresolin et al. (2004) are correct,
the calibration used by Tremonti et al. (2004) may overstate abundances by as
much as 0.5 dex.  This might go some way to explaining why their mass-metallicity
relation implies super-solar abundances for Milky Way mass galaxies.  
Further investigation of the calibration of the strong line method, especially along
the upper branch, is clearly warranted.

\section{Equivalent Width Measurements of $R_{23}$ Lines of McGaugh 1994}

In Table B1 we list the equivalent widths of the H$\beta$, \OII\, and \OIII\ lines of the \HII\ regions of McGaugh (1994) which supplement the current sample of LSB galaxies.  The equivalent widths are used to determine the oxygen abundance, also listed in Table B1, using the EW method described in Section 4.3.  For comparison, the McGaugh (1994) oxygen abundances using the $R_{23}$ method are also listed.

\clearpage

\begin{table}
 \caption{Equivalent Widths and Oxygen Abundances of McGaugh (1994)}
 \begin{tabular}{@{}lcccccc}
 \hline
  Galaxy \& region  &H$\beta$ &\OII\  &\OIII\ &\OIII\ &log(O/H) &log(O/H)  \\
  & &$\lambda$3727 &$\lambda$4959 &$\lambda$5007 &(EW) &($R_{23}$)\\
 \hline 
 F415-3 A1 &36 &89 &38 &137 &$-3.96$ &$-3.94$ \\
 F469-2 A1 &51 &105 &41 &120 &$-4.18$ &$-4.17$ \\
 F469-2 A2 &24 &26 &21 &65 &$-4.41$ &$-4.29$ \\
 F469-2 A3 &39 &44 &73 &240 &$-3.95$ &$-3.98$ \\
 F469-2 A4 &32 &69 &25 &44 &$-4.22$ &$-4.24$ \\
 F530-3 A1 &98 &87 &140 &396 &$-4.22$ &$-4.14$ \\
 F561-1 A3 &22 &55 &18 &54 &$-4.12$ &$-3.86$ \\
 F563-V1 A1 &14 &37 &25 &60 &$-3.90$ &$-3.90$ \\
 F563-V1 A2 &8.5 &7.4 &4.9 &15 &$-4.70$ &$-4.67$ \\
 F563-V2 A1 &12 &16 &6.2 &15 &$-4.55$ &$-4.48$ \\
 F563-V2 A2 &11 &32 &9.9 &33 &$-4.05$ &$-3.60$ \\
 F563-V2 A3 &32 &61 &35 &101 &$-4.13$ &$-3.60$ \\
 F568-6 S1A1 &8.1 &28 &2.9 &8.8 &$-4.13$ &$-3.60$ \\
 F568-6 S3A2 &20 &49 &5.6 &17 &$-3.14$ &$-3.22$ \\
 F577-V1 A3 &29 &109 &30 &94 &$-3.80$ &$-3.60$ \\
 F577-V1 A5 &31 &99 &50 &151 &$-3.71$ &$-3.60$ \\
 F611-1 A1 &47 &66 &54 &160 &$-4.20$ &$-4.14$ \\
 F611-1 A2 &54 &43 &42 &151 &$-4.48$ &$-4.33$ \\
 F746-1 A1 &54 &165 &73 &233 &$-3.78$ &$-3.60$ \\
 F746-1 A2 &13 &78 &14 &44 &$-3.55$ &$-3.60$ \\
 F746-1 A3 &25 &91 &31 &82 &$-3.81$ &$-3.82$ \\
 UGC 1230 A1 &45 &56 &41 &97 &$-4.39$ &$-4.17$ \\
 UGC 1230 A2 &29 &74 &26 &85 &$-4.05$ &$-4.04$ \\
 UGC 1230 A3 &51 &86 &49 &134 &$-4.23$ &$-4.16$ \\
 UGC 1230 A4 &11 &40 &19 &47 &$-3.79$ &$-3.66$ \\
 UGC 1230 A5 &34 &36 &57 &161 &$-4.13*$ &$-4.19*$ \\
 UGC 5709 S1A1 &2.7 &14 &1.5 &4.4 &$-3.22$ &$-3.02$ \\
 UGC 5709 S1A3 &19 &39 &12 &35 &$-3.20$ &$-3.21$ \\
 UGC 6151 S1A1 &48 &118 &48 &148 &$-4.02$ &$-3.65$ \\
 UGC 6151 S1A2 &34 &94 &37 &110 &$-3.96$ &$-3.60$ \\
 UGC 6151 S3A1 &10 &118 &3.6 &11 &$-3.06$ &$-3.60$ \\
 UGC 6151 S3A2 &56 &162 &14 &43 &$-4.13$ &$-4.08$ \\
 UGC 9024 S1A1 &61 &150 &98 &292 &$-3.81$ &$-3.70$ \\
 UGC 9024 S2A1 &55 &164 &58 &173 &$-3.91*$ &$-3.95*$ \\
 UGC 12695 S1A2 &45 &64 &67 &202 &$-4.07$ &$-4.02$ \\
 UGC 12695 S2A1 &92 &70 &169 &518 &$-4.07$ &$-4.14$ \\
 UGC 12695 S2A2 &61 &66 &76 &223 &$-4.25$ &$-4.11$ \\
 UGC 12695 S2A3 &52 &53 &85 &257 &$-4.10$ &$-4.01$ \\
 \hline
 \end{tabular}
\medskip

Equivalent widths in \AA. \\
$^{*}$Ambiguous branch determination; lower branch abundance is listed here.\\
\end{table}

\label{lastpage}

\end{document}